# Peak effect of flux pinning in $Sc_5Ir_4Si_{10}$ superconductors

Y. Sun,[1] Z. X. Shi,[1,2,a)] D. M. Gu,[1] M. Miura,[2] and T. Tamegai[2]
[1]*Department of Physics, Southeast University, Nanjing 211189, China*
[2]*Department of Applied Physics, The University of Tokyo, Hongo, Bunkyo-ku, Tokyo 113-8656, Japan*

Magnetic hysteresis loops (MHLs) have been comparatively measured on both textured and single crystalline $Sc_5Ir_4Si_{10}$ superconductors. Critical current densities and flux pinning forces are calculated from MHLs by Bean model. Three kinds of peaks of the flux pinning force are found at low fields near zero, intermediated fields, and high fields near the upper critical field, respectively. The characters and origins of these peaks are studied in detail. © *2010 American Institute of Physics*. [doi:10.1063/1.3486467]

## I. INTRODUCTION

The $R_5T_4X_{10}$ ($R$=Sc, Y, and Lu; $T$=Co, Rh, Ir, and Os; $X$=Si and Ge) series, first reported in 1980,[1] are ternary compounds with the $Sc_5Co_4Si_{10}$ (P4/mbm)-type structure. The $T$ and $X$ atoms form planar networks of pentagons and hexagons in the basal plane of the primitive tetragonal cell, and these polygons are stacked along the $c$-axis and are connected via $T$-$X$-$T$ zigzag chains. Compounds belonging to the $R_5Ir_4Si_{10}$ family exhibit a variety of ground states depending on the nature of the rare-earth element $R$.[2] Most of them show coexistence of superconductivity and magnetism or charge-density wave.[2,3] Among these compounds, $Sc_5Ir_4Si_{10}$ shows superconductivity with the highest $T_c$ of 8.5 K.[1] However, very few efforts have been made to study the superconducting properties of $Sc_5Ir_4Si_{10}$, especially the flux pinning properties of $Sc_5Ir_4Si_{10}$. In this paper, we have studied the flux pinning behavior of textured and single crystalline $Sc_5Ir_4Si_{10}$ superconductors comparatively.

## II. EXPERIMENTAL

Buttons of $Sc_5Ir_4Si_{10}$ were prepared by arc melting 5:4:10 stoichiometric mixtures of Sc, Ir, and Si in an Ar environment. By cutting the buttons along a certain direction, textured polycrystalline samples were obtained. Single crystals of $Sc_5Ir_4Si_{10}$ used in this work were grown by the floating-zone technique. The ingots of $Sc_5Ir_4Si_{10}$ were first prepared by arc-melting also, then, used to prepare the feeding and seed rods for the floating-zone method. The single crystals were grown with a growth rate of 2 mm/h in an Ar atmosphere with a flow rate of 1 l/min. Two rods were rotated in the opposite directions at 20 rpm. The sizes of the obtained textured and single crystalline samples are about $0.14 \times 0.2 \times 0.06$ $cm^3$ and $0.15 \times 0.1 \times 0.05$ $cm^3$, respectively. Details of the crystal structure were characterized by x-ray diffraction (XRD). The microstructure and element composition of the sample were examined by scanning electron microscopy (SEM) and electron probe microanalyzer (EPMA). Resistivity was measured by physical property measurement system. Magnetization as a function of temperature and magnetic field was measured using a SQUID magnetometer (MPMS).

## III. RESULTS AND DISCUSSION

### A. Characterization of sample

The XRD pattern for the polycrystalline sample is shown in Fig. 1. It can be observed that there are only a few characteristic peaks, of which the highest two peaks correspond to that of (2, 2, 0) and (4, 4, 0), which means that the large surface of this sample is (1, 1, 0) plane and the $c$-axis is parallel to the large surface. According to the Bragg diffraction equation, $2d \sin \theta = n\lambda$, the space between these parallel lattice planes, $d$, was calculated as 8.691 Å and 8.701 Å, respectively. Based on the value of $d$, the lattice parameter $a$ ($b$) was calculated as 12.3 Å equals to the theoretical value. These results manifest that the polycrystalline sample is textured. Superconducting transition temperature $T_c$ of the textured sample was measured by the standard four-probe technique, with the excitation current of 1.0 mA, corresponding to a current density of about 0.19 $A/cm^2$. As shown in Fig. 2, $T_c$ is about 8.3 K and the transition width is about 0.1 K. The very sharp superconducting transition manifests that the sample has a single domain phase.

Zero-field-cooled and field-cooled magnetizations have been measured on both single crystal and textured polycrystalline samples at various fields, as shown in Fig. 3.

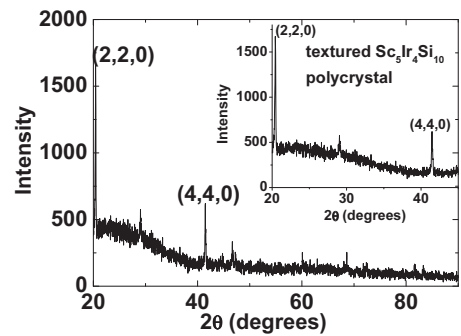

FIG. 1. XRD pattern of $Sc_5Ir_4Si_{10}$ textured sample. Inset is the enlarged part of peak (2, 2, 0) and (4, 4, 0).

a)Electronic mail: zxshi@seu.edu.cn.

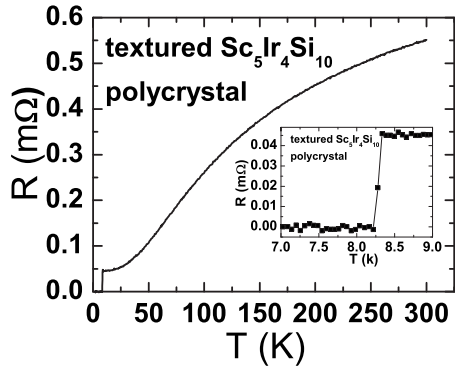

FIG. 2. Temperature dependence of resistance of textured $Sc_5Ir_4Si_{10}$. Inset is the enlarged part of superconducting transition.

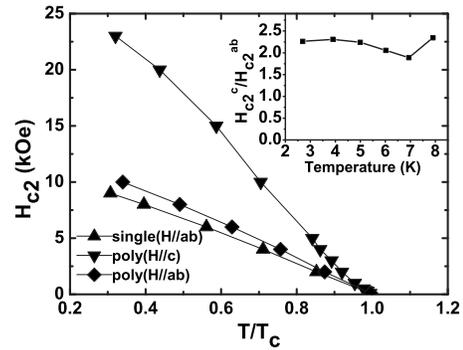

FIG. 4. Temperature dependence of upper critical fields with applied field parallel to *c*-axis and *ab* plane. Inset is the anisotropy of upper critical field of textured sample.

Temperature dependence of the upper critical fields, which were determined by the onset of the diamagnetism, is shown in Fig. 4. The polycrystalline sample has higher $H_{c2}$ possibly due to the microstructural defects. And the anisotropy of upper critical field of the textured polycrystalline sample is about 2.3, as shown in the inset of Fig. 4, which is in agreement with the result of the single crystal sample as reported in Ref. 4. The large anisotropy of $H_{c2}$ indicates that the polycrystalline sample is highly oriented.

### B. Magnetization hysteresis loop and critical current density

Magnetization hysteresis loops (MHLs) were measured at various temperatures with applied field parallel to *c*-axis and *ab* plane on both the single crystalline and textured samples. As shown in Fig. 5, the MHLs of the single crystal sample are asymmetric and show a classic Bean–Livingston (BL) shape. For the field increasing branches, the flux entry was frustrated and magnetizations at various temperatures are quite different; for the field decreasing branches, the flux floods out with negligible barriers and the magnetizations at different temperatures collapse into one curve near zero. However, for the polycrystalline sample, the MHLs are much fatter and almost symmetric about the x-axis, which may be caused by bulk pinning centers and other surface defects rather than BL barriers. Another characteristic of MHLs is the double-peak effect with applied field parallel to *ab* plane, especially for single crystalline sample. The first peak is at fields near zero, while the second is at higher fields near $H_{c2}$. Usually, the first peak is caused by the field dependence of critical current density. The second peak may be caused by the softening of flux lattice at high fields near $H_{c2}$ so that the flux lines can match the pinning centers better. If the closed points of the MHLs were defined as the irreversibility field $H_{irr}$, it is obvious that the $H_{irr}$ in *c*-axis direction is much higher than that in *ab* plane due to the anisotropy of upper critical field.

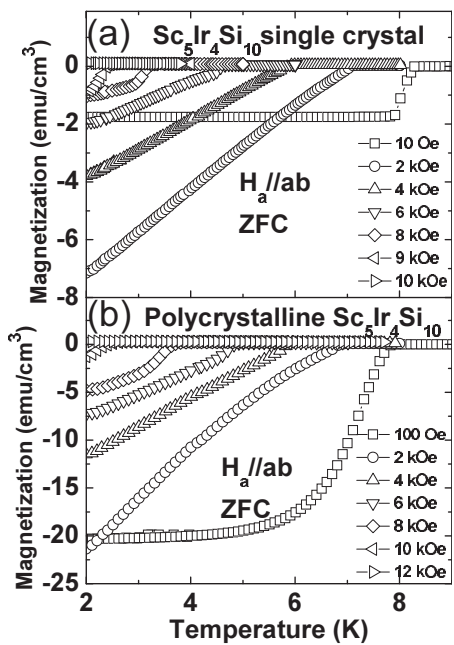

FIG. 3. Temperature dependence of magnetization of single crystal and textured $Sc_5Ir_4Si_{10}$ sample with applied field parallel to *ab* plane.

Critical current densities were calculated by the Bean model.[5,6] As shown in Fig. 5, critical current densities are smaller than $3 \times 10^4$ A/cm$^2$ for polycrystalline sample and smaller than $10^3$ A/cm$^2$ for single crystalline sample. Though $J_c$ of the polycrystalline sample is about one order higher than $J_c$ of the single crystalline sample possibly due to the pinning at the grain boundaries, both the samples are among weak pinning system. The critical current densities of both the single crystal and polycrystalline samples are one order lower than those of $MgB_2$ superconductors,[7,8] respectively. For single crystal sample, there are peaks in the field dependence of $J_c$ at high fields near $H_{c2}$, which are relative to the second peaks in MHLs. The field dependence of $J_c$ has three stages: rapid decrease in lower fields near zero field; exponential field dependence at intermediated fields; peaks and sudden decrease at higher fields near $H_{c2}$. If we determine the irreversibility field by $J_c$, the $H_{irr}$ in *ab* plane are almost the same for both polycrystalline and single crystal samples, though the $J_c$ of the polycrystalline sample is about one order higher than $J_c$ of the single crystal sample. Both $J_c$ and MHLs are determined by flux pinning, and all their characteristics can be explained by pinning mechanism of these samples in Sec. III C.

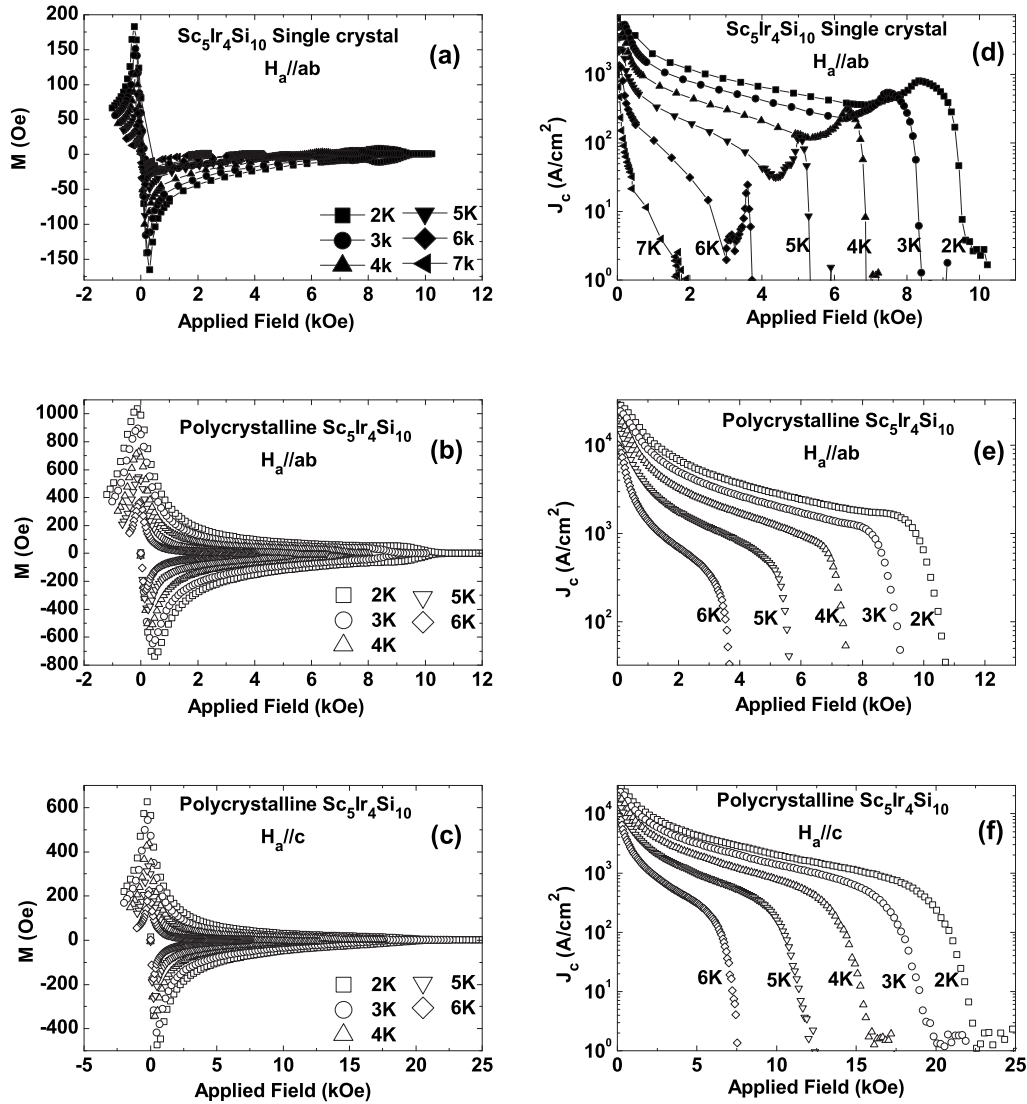

FIG. 5. MHLs and critical current density of textured and single crystal $Sc_5Ir_4Si_{10}$ samples with applied field parallel to *c*-axis and *ab* plane, respectively.

## C. Flux pinning behavior and mechanism

To investigate the pinning mechanism in $Sc_5Ir_4Si_{10}$ superconductors, pinning force densities were calculated from Fig. 5. The field dependence of pinning force is shown in Fig. 6. The scaling behavior of flux pinning force was also investigated by normalizing the applied field and pinning force as $h=H_a/H_{c2}$ and $f_p=F_p/F_{p\,max}$. The scaling behavior is not good due to the influence of pinning force peaks, especially for the single crystalline sample. There are three kinds of peaks: narrow peaks at lower fields near zero; broad peaks at intermediated fields; narrow peaks at higher fields near $H_{c2}$. To study the origin of these peaks, the field dependence of pinning force of the single crystalline and polycrystalline samples is compared in Fig. 6. The pinning mechanisms are discussed as follows.

As shown in Fig. 6, the low-field peaks of flux pinning force are more obvious for the single crystal sample, especially at higher temperatures. These peaks are resulted from surface barriers. First, as pointed out in Sec. III B, the asymmetric MHLs of the single crystal sample manifests that BL surface barriers dominate the MHLs at lower fields. Thus, the low-field peaks of pinning force correspond to surface pinning, which is similar to the situation of untwinned $YBa_2Cu_3O_{7-\delta}$ single crystal.[9] Second, bulk pinning is very weak in clean single crystal samples, especially at higher temperatures. In this case, the effect of surface barriers can be separated from the contribution of bulk pinning. Therefore, low-field peaks of the pinning force can be seen easily in the single crystal sample at higher temperatures. This is an additional evidence for surface barriers. As for the polycrystalline sample, only a small low-field peak can be seen at 6 K with $H//ab$ for its weaker bulk flux pinning. No other low-field peak can be observed for two reasons: (1) Strong bulk pinning surpasses the surface pinning, and the low-field peak is submerged; (2) surface defects of the polycrystalline sample act as pinning centers (no BL barrier) against both the flux entry and flux exit. It is difficult to discriminate surface pinning from bulk pinning.

The peaks at intermediate fields are commonly caused the bulk pinning, which is strong in the polycrystalline sample and weak in the single crystalline sample. As shown in Fig. 6, the pinning force of the polycrystalline sample is about one order larger than that of the single crystal sample. The second peaks are so broad that the pinning forces change smoothly at intermediate fields. On the one hand, the peaks are broadened by both the low-field peaks and the high-field peaks. The second-peak positions shift slowly to lower fields with increasing temperature due to the influence of the surface pinning. Especially for polycrystalline sample with applied field H//surface, the second-peak positions shift from 0.3 to 0.1. With the applied field parallel to the sample surface (parallel to *c*-axis), more surface defects act as effective pinning centers, thus the surface pinning is so strong that the first peak and the second peak merge together, as shown in Fig. 6(f). Only with applied field perpendicular to sample surface (parallel to *ab* plane), the first peak and the second peak can be separated as shown in Figs. 6(d) and 6(e).

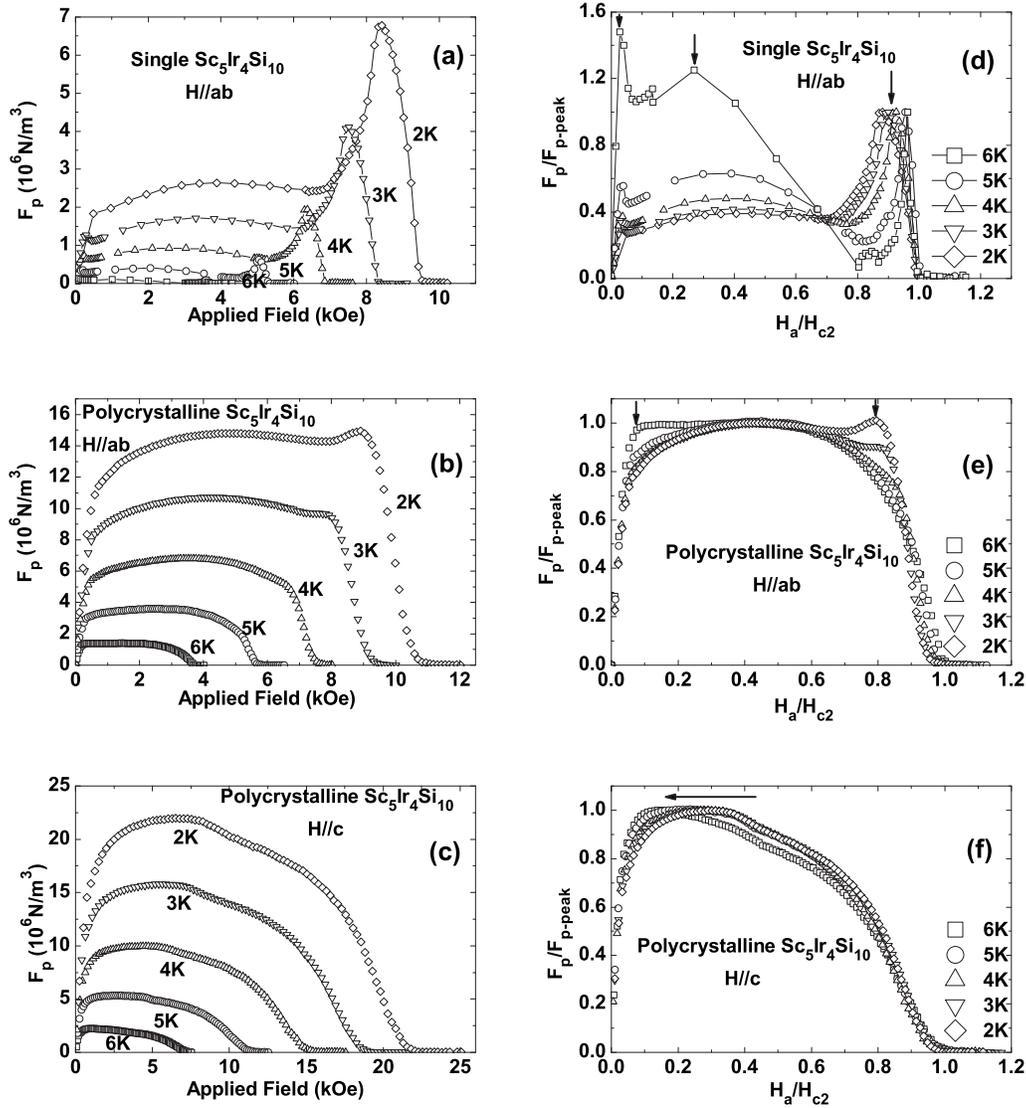

FIG. 6. Field dependence of flux pinning force density calculated from Fig. 5 and scaling behavior of pinning force of textured and single crystal $Sc_5Ir_4Si_{10}$ samples.

On the other hand, the broad peaks suggest that more than one pinning mechanism may be operative in these samples. According to Dew–Hughes model,[10] precipitates together with other crystal defects could give rise to a mixture of magnetic and core pinning. Their combined effects result in broad peaks between 0.2 and 0.67. To check the origin of the pinning behavior, microstructural and chemical analyses were performed by SEM. As shown in Fig. 7, small spots were observed in both the single crystalline and polycrystalline $Sc_5Ir_4Si_{10}$ multiphase. There are dense white spots in the polycrystalline sample and sparse black spots in the single crystal sample. The element composition of these spots was determined by EPMA as shown in Table I. These results manifest that both the single crystalline and polycrystalline samples contain a dispersion of small, nonsuperconducting precipitates. The average size of the small spots is about several micrometer, which is much larger than the coherence length $\xi_{ab} \approx 12$ nm and $\xi_c \approx 23$ nm estimated from the $H_{c2}(0)$. The precipitate size and their distance are also larger than the field penetration depth $\lambda$ (84 nm in *c*-axis and 210 nm in *ab* plane).[4] Thus, these precipitates can be volume pinning centers for both the magnetic and core pinning. Since the peak position is at about $h_p \approx 0.4$ as shown in Figs.

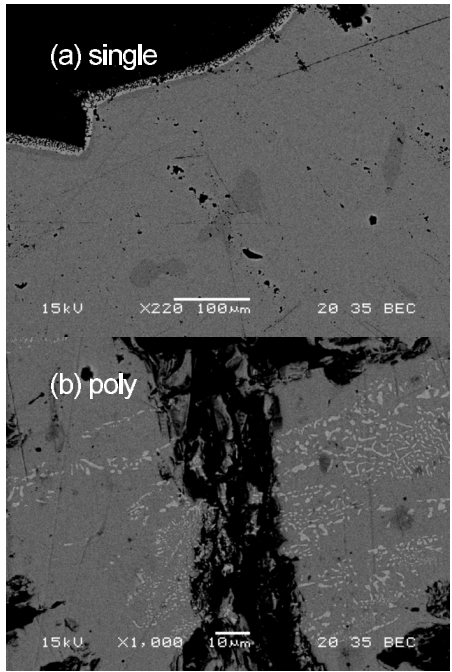

FIG. 7. SEM images of textured and single crystal $Sc_5Ir_4Si_{10}$ samples.

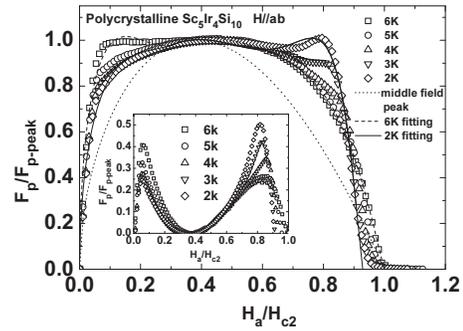

FIG. 8. Scaling behavior of pinning force of textured sample for $H//ab$. The dot line represents the calculated middle field peak including the contribution from both volume magnetic normal pinning and volume core pinning. Inset is the low-field and high-field peaks obtained by subtracting the middle field peak. The solid and dash line separately manifest the fitting results of reduced pinning force at 2 and 6 K.

6(f) and 6(e), the possible pinning mechanism could be magnetic normal pinning or core $\Delta\kappa$ pinning by volume pinning centers.

To confirm the discussion above, Fig. 6(e) was redrawn to fit the experimental data. As shown in Fig. 8, the dotted line is the calculated pinning force including the contribution from both the volume magnetic normal pinning and the volume core $\Delta\kappa$ pinning as $f_{M.peak}=(2/3)f_{mag}+(1/3)f_{core}$, where $f_{mag}=f_p(0)h^{1/2}(1-h)$ and $f_{core}=f_p(0)h(1-h)$. Though the peak position is the same as experimental data, the theoretical curve could not fit the experimental data due to the influence of peaks in lower fields and higher fields. By subtracting $f_{M.peak}$ from the total pinning force, we got the contribution from the low-field peak $f_{L.peak}$, and the high-field peak $f_{H.peak}$, as shown in the inset of Fig. 8. Both the $f_{L.peak}$ and the $f_{H.peak}$ are similar to the low-field and high-field peaks of the single crystal sample, such as the high-field peak moves to higher reduced field with increasing temperature, while the low-field peak does not change peak position. The total pinning force can be written as $f_p=f_{L.peak}+f_{M.peak}+f_{H.peak}$. For clarity, the typical fitting curves of 2 and 6 K are shown as the solid and dashed line in Fig. 8.

Finally, the $F_p$ peaks at higher fields near $H_{c2}$ can be caused by several factors. In our case, the peak appears eas-ily with applied field parallel to $ab$ plane at lower temperatures. It seems related to the softening of flux lattice and the anisotropy of superconductivity. $Sc_5Ir_4Si_{10}$ shows quasi-one-dimensional superconductivity as reported in Ref. 11. Superconductivity in $c$-axis is much stronger than that in $ab$ plane so that the coherence in $ab$ plane is shorter. When the field is applied parallel to the ab plane, flux lattice will soften more easily at higher fields. The softened flux lattice can match the pinning centers better and result in peak effects.

However, compared with the polycrystalline sample, the single crystalline sample shows poor scaling behavior of flux pinning force. The high-field-peak position shifts to lower reduced field h at lower temperatures, as shown in Fig. 6(d). There are some reasons such as the influence of elementary pinning force threshold and the influence of Pauli paramagnetism can result in the breakdown of simple scaling. In our case, the single crystalline sample has weaker pinning centers. The reason is more like the paramagnetism rather than the elementary pinning force threshold, since the effect of paramagnetism is stronger in the high field region.[12] According to Werthamer-Helfand-Hohenberg-Maki (WHHM) model,[13,14] paramagnetic limitation causes the temperature dependence of Maki parameter $\kappa_2$. The shift in the value of $\kappa_2$ will result in considerable shift in the temperature of upper critical field $H_{c2}(t)$ and pinning force density, especially at higher fields. The reduced pinning curves will spread apart. Full scaling can be achieved if the appropriated temperature dependence of $\kappa_2$ is introduced at higher fields. As shown in Fig. 9, the modification of $F_p/F_{p.\max}$ in two regimes represented by $h=0.2$ and $h=0.8$ is carried out. Scaling was restored for $f_p \sim h$ curves at 2, 3, and 4 K, as shown in Fig. 9(b). The direct evidence of paramagnetism influence on $Sc_5Ir_4Si_{10}$ superconductivity will be discussed in our next paper.

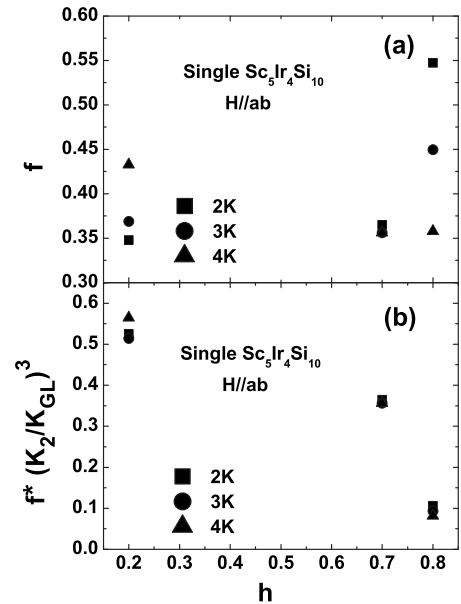

FIG. 9. $f_p \sim h$ curves of single crystal $Sc_5Ir_4Si_{10}$ sample at 2, 3, and 4 K before (a) and after (b) modification.

TABLE I. Element composition of textured and single crystal $Sc_5Ir_4Si_{10}$ samples.

| Element | Stoichiometric ratio (%) | Polycrystalline | | Single crystal | |
| --- | --- | --- | --- | --- | --- |
| | | Main body (%) | White dot (%) | Main body | Black dot |
| Sc | 26.3 | 27.5 | 10.0 | 26.7 | 18.1 |
| Ir | 21.1 | 22.0 | 28.4 | 22.5 | 21.5 |
| Si | 52.6 | 50.6 | 61.7 | 50.9 | 60.4 |

## IV. CONCLUSIONS

$Sc_5Ir_4Si_{10}$ superconductors are clean systems with weak flux pinning. Surface and bulk pinning resulted in zero-field and intermediate field peaks, respectively. Flux lattice softening caused the high-field peaks of flux pinning force and critical current density, especially for the single crystalline sample with field parallel to *ab* plane. Paramagnetic effect could be responsible for the non-scaling behavior of the high field peaks in the single crystal sample.

## ACKNOWLEDGMENTS

This work was supported by the Ministry of Education of the People's Republic of China (Grant No. NCET-05-0461), the Jiangsu Industry Support Project (Grant No. BE2009053), and a grant from the National Science Foundation of China (Grant No. 10904013).